\documentstyle[twocolumn,aps,prd]{revtex}

\begin{document}

\draft

\title{An Efficient Method for Fully Relativistic Simulations \\
of Coalescing Binary Neutron Stars}

\author{Walter Landry}

\address{Physics Dept., University of Utah, SLC, UT 84112 \\
and Center for Radiophysics and Space Research, Cornell University, Ithaca,
NY 14853}

\author{Saul A. Teukolsky}

\address{Center for Radiophysics and Space Research, Cornell University, Ithaca, NY
14853}

\maketitle
\begin{abstract}
The merger of two neutron stars has been proposed as a source of gamma-ray bursts,
\( r \)-process elements, and detectable gravitational waves. Extracting information
from observations of these phenomena
requires fully relativistic simulations. Unfortunately, the
only demonstrated method for stably evolving neutron stars requires solving
elliptic equations at each time step, adding substantially to the computational
resources required. In this paper we present a simpler, more efficient method.
The key insight is in how we apply numerical diffusion. We perform a number
of tests to validate the method and our implementation. We also carry
out a very rough
simulation of coalescence and extraction of the gravitational waves
to show that the method is viable if realistic initial data are provided.
\end{abstract}


\section{Introduction}

After the discovery of the first binary neutron star system 1913+16, it was
quickly realized that the orbit was decaying and the two stars would collide
within a Hubble time. Because of the enormous available gravitational binding
energy, neutron star-neutron star mergers like this became a popular mechanism
for \( \gamma  \)-ray bursts (e.g., \cite{Narayan}). The merger could also
eject some neutron star material, which led some to propose that the subsequent
rapid decompression of the nuclear density material could create \( r \)-process
elements \cite{Symbalisty,Eichler in Nature,sph r-process}. 

However, whether or not neutron star - neutron star mergers can explain these
phenomena is still uncertain. What is reasonably certain is that, if general
relativity is correct, the coalescence will emit a gravitational wave signal
visible at cosmological distances to detectors being built or planned today
\cite{Abramovici}. By comparing the signals at coalescence to theoretical predictions,
we may be able to learn more about the neutron stars themselves: their mass,
radius, and internal structure. Because neutron stars are such dense, relativistic
objects, such a comparison could also lead to new understandings in the physics
at nuclear densities as well as the first strong field test of general relativity.

When the stars collide, shocks form, black holes appear, and in general the
physics is very complicated. This requires a fully dynamical scheme to evolve
the gravitational and matter fields. There have been a number of Newtonian and
post-Newtonian simulations (see \cite{Nakamura} and references
therein). This approach
expands the gravitational field equations around the Newtonian limit, using
the ratio of the mass to radius, \( M/R \), as a small parameter (We use units
where \( G=c=1 \)). The problem is that the coalescence is highly relativistic.
The quantity \( 2M/R=0.4 \) for the canonical neutron star with \( M=1.4M_{\odot } \),
\( R=10\mathrm{km} \). It goes up to \( 2M/R=1 \) when the black hole forms,
a process that Newtonian and post-Newtonian simulations cannot even capture.

To treat this case, there have been some attempts to do a fully relativistic
simulation, but with limited success. General relativity is a very complicated
theory. Adding in the details of a numerical implementation leads to a multitude
of things that can go wrong. As a result, self-consistent hydrodynamics+relativity
simulations have had their greatest successes in only one or two dimensions
\cite{Stark,Evans,Schinder,Mezzacappa,Shapiro & Teukolsky,Abrahams}. These calculations
have to evolve both the hydrodynamics (the matter) and the gravitational fields
(the metric). In addition, general relativity allows the freedom of choosing
coordinates. The calculations have all been carried out with variations of the
formalism developed by Arnowitt, Deser and Misner (ADM) \cite{ADM}. These lower
dimensional calculations used a few tricks and applied brute force to manage
long evolutions. In three dimensions, these techniques are either unavailable
or impractical. 

The neutron star problem differs from evolving black holes alone. In one sense,
black holes are easier because there are no difficulties associated with the
hydrodynamics or the (physically) sharp matter distribution at the surface.
On the other hand, event horizons and singularities pose significant problems
for numerical implementations in the black hole case, which have yet to be solved
in three dimensions.

Undaunted, Nakamura \emph{et. al.} \cite{Nakamura} made some pioneering calculations
of coalescing neutron stars. They came up against the same problem that has
plagued numerical relativity until recently: growing instabilities that quickly
crash the code. Marronetti \emph{et. al.} \cite{Wilson} circumvented this problem
by using a simplified metric. Unfortunately, using the simplified metric precludes
an accurate determination of the gravitational waveform.

Font \emph{et. al.} \cite{Miller et. al.} have constructed a 3D code that has
made some short calculations \cite{Miller conjecture}. They have managed long
\emph{hydrodynamic} evolutions of single neutron stars (i.e., they kept the
gravitational fields fixed to their analytic values), but a long term, self-consistent,
stable evolution remained elusive.

Recently, Baumgarte \emph{et. al.} \cite{Baumgarte} modified the original ADM
equations, trying to remove pathologies from the equations which might lead
to instabilities. The tradeoff is that the implementation is a little more complex.
They applied this formalism to a number of spacetimes \cite{Thomas star}, including
a static neutron star. In contrast to the work \emph{}in \emph{}\cite{Miller et. al.},
they evolved the gravitational fields, but kept the hydrodynamic variables frozen
to the analytic solution. They were rewarded with long term stability for this
pseudo-dynamic problem. 

Baumgarte \emph{et. al.} \cite{Baumgarte self-consistent} applied this method
to a fully self-consistent hydrodynamic evolution of a neutron star. They were
able to evolve the star for several dynamical times, but the numerical errors
eventually caused the star to collapse upon itself. It is also not clear whether
their choice of coordinates will work with two coalescing stars. The motion
of two stars around the center of mass could drag the coordinates along. This
twisting around the center could lead to coordinate singularities, ruining an
otherwise physically valid simulation.

To alleviate this twisting, Shibata \cite{Shibata} adopted a more robust coordinate
condition (technically, approximate minimal distortion), and got a long term,
stable, self-consistent, 3D relativistic simulation of coalescence. Unfortunately,
the coordinate condition is computationally expensive (it requires solving elliptic
equations at every time step), and will significantly slow down any simulation.

This paper presents a simpler, more efficient method than \cite{Shibata} for
evolving relativistic stars. We take elements from many different previous investigations,
add in a few small new contributions, and combine them into a stable code. We
use the original ADM formalism to evolve the metric. Although numerical relativists
seem to shun them, we choose fully harmonic coordinates because they may help
alleviate coordinate pathologies. There have been some attempts to use harmonic
\emph{slicing} (e.g., \cite{Bona et. al. 1998}), but not the full harmonic
gauge. We use the relatively new and sophisticated high resolution shock capturing
method for relativistic hydrodynamics developed \emph{}in \emph{}\cite{Miller et. al.}.
For the outer boundaries, we adopt the condition from \cite{Baumgarte} that
allows gravitational waves to propagate off the grid. Our new contribution is
how we add in a small numerical diffusion to stabilize the simulation. The usual
way to add it in diffuses mostly along coordinate axes (\( x \), \( y \),
\( z \)). We use a new scheme that also diffuses along diagonals (\( x=y \),
\( x=-y \), \( x=z \), \( x=-z \), \( y=z \), \( y=-z \)). This small variation
allows long term, accurate evolutions.

We then present a number of tests: short term tests to validate our implementation
of the equations, and long term tests to validate the whole approach. Then we
proceed to compute a very unrealistic simulation of coalescing binary neutron
stars to demonstrate that the method will work. Finally, we describe what remains
to be done to turn this into a realistic simulation. We expect that reasonably
accurate simulations will become available in the next few years, although they
may require the largest supercomputers.

\section{Methods}

\subsection{Metric Evolution}

To evolve the metric, we use the standard decomposition of Arnowitt, Deser and
Misner (ADM) \cite{ADM}. The line element then takes the form 
\begin{equation}
\label{first}
ds^{2}=g_{\mu \nu }dx^{\mu }dx^{\nu }=-\alpha ^{2}dt^{2}+\gamma _{ij}(dx^{i}+\beta ^{i}dt)(dx^{j}+\beta ^{j}dt).
\end{equation}
 The lapse and shift functions \( \alpha  \) and \( \beta _{i} \) embody the
gauge freedom of general relativity (i.e., the freedom to choose arbitrary coordinates),
and so can be chosen without restriction. The embedding of the \( t=\mathrm{constant} \)
slices in the spacetime is described geometrically by the extrinsic curvature
\( K_{ij} \), which is defined by the equation 
\begin{equation}
\label{g dot}
\frac{\partial \gamma _{ij}}{\partial t}=-2\alpha K_{ij}+\nabla _{i}\beta _{j}+\nabla _{j}\beta _{i},
\end{equation}
 where \( \nabla _{i} \) denotes a covariant derivative with respect to the
three dimensional metric \( \gamma _{ij} \). We have written this equation
in the form of an evolution equation for \( \gamma _{ij} \). Using Einstein's
equations one can derive the evolution equation for \( K_{ij} \):

\begin{eqnarray}
\frac{\partial K_{ij}}{\partial t}=-\nabla _{i}\nabla _{j}\alpha +K_{lj}\nabla _{i}\beta ^{l}+K_{il}\nabla _{j}\beta ^{l}+\beta ^{l}\nabla _{l}K_{ij} &  & \nonumber \\
+\alpha \left[ R_{ij}-2K_{il}K^{l}_{j}+KK_{ij}-S_{ij}-\frac{1}{2}\gamma _{ij}(\rho -S)\right] . & \label{K dot} 
\end{eqnarray}

This in turn introduces the Ricci tensor \( R_{ij} \) and the matter terms
\( \rho  \) and \( S \). \( R_{ij} \) is the three-dimensional Ricci tensor
associated with the metric \( \gamma _{ij} \). The most common way to write
it involves first constructing the connection coefficients \( \Gamma ^{i}_{jk} \)
with the equation 
\[
\Gamma ^{i}_{jk}=\frac{\gamma ^{il}}{2}(\gamma _{lj,k}+\gamma _{lk,j}+\gamma _{jk,l}),\]
 where commas denote partial derivatives. Then we could use the standard textbook
formula to construct \( R_{ij} \). Unfortunately, this involves taking derivatives
of \( \Gamma ^{i}_{jk} \). In taking a derivative of a derivative numerically,
there is no simple way to keep the error in the derivatives second order convergent
and have the finite difference stencil include only nearest neighbor couplings.
Second order accuracy requires knowledge of points two grid points away. This
complicates boundary conditions, because we would have to apply them to a layer
of points two points deep instead of one point deep. So, we write the Ricci
tensor in an alternative form as 
\begin{eqnarray}
R_{ij}=\frac{1}{2}\gamma ^{kl} & \left[ \gamma _{kj,il}+\gamma _{il,kj}-\gamma _{kl,ij}-\gamma _{ij,kl}\right.  & \nonumber \\
 & \left. +2\left( \Gamma ^{m}_{il}\Gamma _{mkj}-\Gamma ^{m}_{ij}\Gamma _{mkl}\right) \right] . & \label{Ricci} 
\end{eqnarray}
 Now, we can take centered, second-order differences of the metric \( \gamma _{ij} \),
for which we only need to use the nearest neighbors.

\( \rho  \) and \( S \) are projections of the four-dimensional stress energy
tensor \( T^{\mu \nu } \) onto the three-dimensional spacetime described by
\( \gamma _{ij} \). They are defined in terms of the 4-vector normal to the
three dimensional slice 
\[
n_{\mu }=(-\alpha ,0,0,0).\]
 We then define three projections, \( \rho  \), \( J^{i} \), and \( S_{ij} \)
as
\begin{eqnarray*}
\rho  & = & 8\pi n_{\mu }n_{\nu }T^{\mu \nu }=8\pi \alpha ^{2}T^{tt},\\
J^{i} & = & -8\pi n_{\mu }\gamma ^{i}_{j}T^{\mu j},\\
S_{ij} & = & 8\pi \gamma _{ik}\gamma _{jl}T^{kl},
\end{eqnarray*}
 and
\[
S=\gamma ^{ij}S_{ij}.\]

\subsection{Constraints}

The evolution equations (eqs. \ref{g dot} and \ref{K dot}) involve only 6
of the 10 Einstein equations. The remaining equations are the energy, or Hamiltonian,
constraint 
\begin{equation}
\label{Energy}
R+K^{2}-K_{ij}K^{ij}=2\rho ,
\end{equation}
 and the three momentum constraints 
\begin{equation}
\label{Momentum}
\nabla _{j}\left( K^{ij}-\gamma ^{ij}K\right) =J^{i}.
\end{equation}

If we start with data that satisfies these constraints and evolve them with
eqs. \ref{g dot} and \ref{K dot}, then we are guaranteed to evolve to a spacetime
that still satisfies the constraints. (Note that it doesn't matter how \( \alpha  \)
and \( \beta ^{i} \) are evolved. Their evolution is determined by the gauge,
and eqs. \ref{Energy} and \ref{Momentum} are true in any gauge.)

This guarantee is analytic, but errors in a numerical evolution can lead to
a (\( \gamma _{ij} \), \( K_{ij} \)) pair that does not exactly satisfy the
constraints. Accordingly, we generally use the constraints as a check on the
accuracy of the code. However, in order to get self-consistent initial data
for two neutron stars, we need to explicitly enforce the constraints. In doing
this, an ambiguity immediately presents itself. Eqs. \ref{Energy} and \ref{Momentum}
do not declare what combination of the twelve components of \( \gamma _{ij} \)
and \( K_{ij} \) should be constrained by the four constraints. A natural way,
but by no means the only way, is to decompose \( \gamma _{ij} \) and \( K_{ij} \)
with the York decomposition \cite{York}. We decompose the metric by defining
a conformal metric \( \widetilde{\gamma }_{ij}=\gamma _{ij}\psi ^{-4} \), where
\( \det \widetilde{\gamma }_{ij}=1 \). The inverse of \( \widetilde{\gamma }_{ij} \)
is then \( \widetilde{\gamma }^{ij}=\gamma ^{ij}\psi ^{4} \). We then apply
the energy constraint to \( \psi  \).

Decomposing the extrinsic curvature is a little more complicated. We decompose
the extrinsic curvature into traced and trace-free parts
\[
K^{ij}=\psi ^{-10}\left( \widetilde{A}^{ij}+\left( lX\right) ^{ij}\right) +\frac{1}{3}\psi ^{-4}\widetilde{\gamma }^{ik}\mathrm{Tr}K,\]
 where
\[
\left( lX\right) ^{ij}=\widetilde{\nabla }^{i}X^{j}+\widetilde{\nabla }^{j}X^{i}-\frac{2}{3}\widetilde{\gamma }^{ij}\widetilde{\nabla }_{k}X^{k},\]
 and all covariant derivatives \( \widetilde{\nabla } \) are with respect to
the conformal metric \( \widetilde{\gamma _{ij}} \). This is not a unique decomposition,
since \( \widetilde{A}^{ij} \) and \( (lX)^{ij} \) describe overlapping degrees
of freedom. In practice, we decompose \( K_{ij} \) assuming \( X^{i}=0 \),
solve for a new \( X^{i} \) which makes \( K_{ij} \) solve the constraints,
and recompose \( K_{ij} \).

In terms of these variables, the constraints become 
\begin{eqnarray}
-8\widetilde{\nabla }^{2}\psi  & = & -\widetilde{R}\psi -\frac{2}{3}\left( trK\right) ^{2}\psi ^{5}\nonumber \\
 &  & +\left( \widetilde{A}^{ij}+\left( lX\right) ^{ij}\right) ^{2}\psi ^{-7}+2\rho \psi ^{5},\label{phi constraint} \\
\widetilde{\nabla }^{2}X^{i} & + & \frac{1}{3}\widetilde{\nabla }^{i}\widetilde{\nabla }_{j}X^{j}+\widetilde{R}^{i}_{j}X^{j}\nonumber \\
 & = & J^{i}\psi ^{10}-\widetilde{\nabla }_{j}\widetilde{A}^{ij}+\frac{2}{3}\psi ^{6}\widetilde{\nabla }^{i}trK,\label{X constraint} 
\end{eqnarray}
 where \( \widetilde{R} \) is the Ricci tensor associated with the conformal
metric \( \widetilde{\gamma } \). To solve these equations, we first linearize
them, writing \( \psi  \) as \( \psi _{0}+\delta \psi  \) and \( X^{i} \)
as \( X_{0}^{i}+\delta X^{i} \). We explicitly decouple the linearized equations
by setting \( \delta X^{i}=0 \) in the linearized form of eq. \ref{phi constraint},
and \( \delta \psi =0 \) in the linearized form of eq. \ref{X constraint}.
The linearized equations then become 
\begin{eqnarray}
-8\widetilde{\nabla }^{2}(\psi _{0}+\delta \psi )=-\widetilde{R}(\psi _{0}+\delta \psi )-\frac{2}{3}\left( trK\right) ^{2}\psi _{0}^{4}(\psi _{0}+5\delta \psi ) &  & \nonumber \\
+\left( \widetilde{A}^{ij}+\left( lX_{0}\right) ^{ij}\right) ^{2}\psi _{0}^{-8}(\psi _{0}-7\delta \psi )+2\rho \psi _{0}^{4}(\psi _{0}+5\delta \psi ), &  & \label{phi linear} 
\end{eqnarray}
\begin{eqnarray}
\nabla ^{2}(X_{0}^{i}+\delta X^{i})+\frac{1}{3}\widetilde{\nabla }^{i}\widetilde{\nabla }_{j}(X_{0}^{j}+\delta X)+\widetilde{R}^{i}_{j}(X_{0}^{j}+\delta X^{i}) &  & \nonumber \\
=J^{i}\psi _{0}^{10}-\widetilde{\nabla }_{j}\widetilde{A}^{ij}+\frac{2}{3}\psi _{0}^{6}\widetilde{\nabla }^{i}trK. &  & \label{X linear} 
\end{eqnarray}

Then we solve the linearized equations using multigrid methods \cite{Recipes}.
We use red-black Gauss-Seidel smoothing, with 2 pre- and post-smoothings for
the linearized \( \psi  \) equation, and 5 pre- and post-smoothings for the
linearized \( X^{i} \) equation. On the outer boundaries, we assume that \( \psi =\psi _{\mathrm{original}} \)
and \( X^{i}=X_{\mathrm{original}}^{i}=0 \) (i.e., we freeze the boundaries).
When working on parallel machines, we smooth the solution on each machine individually,
and then synchronize the edges. Then we add \( \delta \psi  \) to \( \psi _{0} \)
and \( \delta X^{i} \) to \( X_{0}^{i} \), recompute the terms involving \( \psi _{0} \)
and \( X^{i}_{0} \) in eqs. \ref{phi linear} and \ref{X linear}, and iterate
until we reach a tolerance of \( 10^{-10} \) in the norm of eqs. \ref{phi linear}
and \ref{X linear}.

The explicit decoupling helps the elliptic solver converge. Decoupling does
not affect the results, since we recompute the non-linear parts of eqs. \ref{phi linear}
and \ref{X linear} after each solution of the linearized equations.

As mentioned before, the constraints are only solved to get self-consistent
initial data. The constraints are not solved during the evolution.

\subsection{Coordinate Evolution}

In general relativity you have the freedom to choose the coordinates however
you wish. This freedom manifests itself in the choice of original coordinates
as well as the evolution equations for the lapse \( \alpha  \) and the shift
\( \beta ^{i} \). It is very easy to make a bad choice of coordinates. For
example, consider a small perturbation in the original coordinates \( x^{\mu } \)
\begin{equation}
\label{gauge perturbation}
x^{\mu }\rightarrow x^{\mu }+\xi ^{\mu }.
\end{equation}

A seemingly benign choice of evolution equations for \( \alpha  \) and \( \beta ^{i} \)
can make the small perturbation grow exponentially, creating coordinate singularities
in an otherwise non-singular evolution. To prevent any bad behavior in \( \xi ^{\mu } \),
we can use harmonic coordinates. These coordinates satisfy 
\begin{equation}
\label{gauge}
\Box x^{\mu }=0.
\end{equation}
 That is, the coordinates obey a wave equation (although with covariant derivatives).
Since eq. \ref{gauge} is linear, and the original coordinates \( x^{\mu } \)
obey eq. \ref{gauge}, we then get the condition on \( \xi ^{\mu } \) 
\begin{equation}
\label{perturbation}
\Box \xi ^{\mu }=0.
\end{equation}
 Thus, \( \xi ^{\mu } \) \emph{should} be wave like, and not exponentially
increasing.

Imposing eq. \ref{gauge} implies evolution equations for the lapse and shift.
In practice, we use the evolution equation for \( g_{tt}=-\alpha ^{2}+\beta ^{i}\beta _{i} \)
instead of the lapse. Then the equations are 
\begin{equation}
\label{g_tt dot}
\frac{\partial g_{tt}}{\partial t}=\left( \gamma ^{ij}\alpha ^{2}-\beta ^{i}\beta ^{j}\right) \left( -\gamma _{ij,t}+2\beta _{i,j}\right) +2\beta ^{i}g_{tt,i}
\end{equation}
 and 
\begin{eqnarray}
\frac{\partial \beta _{k}}{\partial t} & = & 2\beta ^{i}\left( \gamma _{ki,t}-\beta _{i,k}+\beta _{k,i}\right) \nonumber \\
 &  & -\left( \gamma ^{ij}\alpha ^{2}-\beta ^{i}\beta ^{j}\right) \left( \gamma _{ij,k}-2\gamma _{kj,i}\right) +g_{tt,k}.\label{Shift dot} 
\end{eqnarray}

\subsection{Matter Evolution}

Our method is exactly the same as in \cite{Miller et. al.} except for our treatment
of low density regions and our time stepping. We briefly summarize the method
here.

We represent the matter with three variables --- \( D \), \( \tau  \) , and
\( S_{i} \) --- which roughly correspond to the mass density, energy density,
and momentum density. We can then write the evolution equations in the form
\[
\frac{\partial (\mathrm{variable})}{\partial t}+\partial _{i}(\mathrm{flux})^{i}=(\mathrm{source}),\]
 where the only spatial derivatives of the \emph{matter} variables occur in
the \( \partial _{i}(\mathrm{flux})^{i} \) term. Writing the equations in this
form allows us to apply the cornucopia of shock capturing methods developed
over the years for ordinary hydrodynamics. We use a Roe scheme with a standard
min-mod piecewise-linear reconstruction algorithm \cite{Leveque book}. 

The fluxes and sources are written in terms of primitive variables \( \{\rho ,\epsilon ,v_{i}\} \)
which are not simply related to the evolved variables \( \{D,\tau ,S_{i}\} \).
Thus we have to do a costly root finding at each point at each time level to
find \( \{\rho ,\epsilon ,v_{i}\} \) from \( \{D,\tau ,S_{i}\} \). This step
can go awry in low density regions, because numerical errors can conspire to
evolve \( \{D,\tau ,S_{i}\} \) into something that has no physical \( \{\rho ,\epsilon ,v_{i}\} \).
To combat this, we use a variation of the method of \cite{Miller et. al.}.
We create a fake ``atmosphere'' in the empty space around the star by setting
a minimum value of \( D \) of about \( 10^{-9} \) of the initial central density
of the star. Whenever \( D \) is evolved to a value lower than \( 10^{-9} \),
we set \( D \) to \( 10^{-9} \). Also, if \( D \) is lower than \( 10^{-5} \),
we set \( \tau  \) and \( S_{i} \) to zero. Furthermore, if we ever get a
transform from \( \{D,\tau ,S_{i}\} \) to \( \{\rho ,\epsilon ,v_{i}\} \)
that gives unphysical values (e.g., negative energies), we start over, replacing
the definition of \( \tau  \) with the condition for adiabatic flow \( P=k\rho ^{\Gamma } \).
We then compute \( \{\rho ,\epsilon ,v_{i}\} \) and recompute \( \tau  \).
This simple prescription allows long term evolutions of neutron stars and their
surroundings.

Apart from the low density treatment, the only other difference between us and
\cite{Miller et. al.} is the time stepping. We implement a Strang split of
the hydrodynamics \cite{Leveque book}. That is, we evolve for a half time step
as if the evolution equations are only the flux terms of the hydrodynamics.
Then we evolve for a full time time step as if the evolution equations are only
the hydrodynamic source terms and all the metric and coordinate terms. Then
we again evolve for a half time step as if the evolution equations are only
the flux terms. 

When we evolve the fluxes for the first half step, we evolve first in one randomly
chosen direction \( d_{1} \), then in another randomly chosen direction \( d_{2} \),
and then in the third direction \( d_{3} \). For the second half-step, we do
it in the reverse order \( d_{3} \)-\( d_{2} \)-\( d_{1} \). Thus, for each
step we choose from one of six random orders: \( xyz \)-\( zyx \), \( xzy \)-\( yzx \),
\( yxz \)-\( zxy \), \( yzx \)-\( xzy \), \( zxy \)-\( yxz \), and \( zyx \)-\( xyz \). 

We use second-order Runge-Kutta stepping for the matter, metric, and coordinate
equations. That is, when evolving the equation 
\[
\frac{\partial Q}{\partial t}=F(Q)\]
 from \( t \) to \( t+\Delta t \), we first take an intermediate step 
\[
Q_{\mathrm{intermediate}}=Q_{t}+\frac{\Delta t}{2}F(Q_{t}),\]
 and then we take the full step 
\[
Q_{t+\Delta t}=Q_{t}+\Delta t\, F(Q_{\mathrm{intermediate}}).\]
In conjunction with the Strang split, this makes the whole evolution second-order
accurate. We use a Courant factor \( \Delta t/\Delta x=0.25/c \). This scheme
is different from \cite{Miller et. al.}. They evolve the fluxes in the \( x \),
\( y \), and \( z \) direction all at once, and their coupling in time between
hydrodynamics and spacetime is very different.

\subsection{Boundary Conditions}

On the outer boundaries, we use the interpolated Sommerfeld condition of \cite{Baumgarte}
for the metric and coordinate terms. Essentially, we assume that all of the
metric and coordinate variables behave like outgoing, radial waves 
\[
Q(t,r)=\frac{G(\alpha t-(\det \gamma )^{\frac{1}{6}}r)}{r},\]
 where \( Q=\{\gamma _{ij}-\delta _{ij},K_{ij},g_{tt}+1,\beta _{i}\} \). Thus,
we compute the value at the boundary by following the characteristic back to
the previous time step and linearly interpolating the corresponding variable
to that point. We tried just freezing the boundaries (i.e., \( \gamma _{\mathrm{old}}=\gamma _{\mathrm{new}} \)),
but that seemed to reflect and amplify waves, leading to an instability.

For the matter terms, we use ``flat'' boundaries. That is,
values at points on the boundary
are set equal to those at the points just inside the boundary. Although this condition
seems crude, it works rather well at propagating blobs of matter off of the
computational grid \cite{Leveque}.

\subsection{Numerical Diffusion}

Evolutions using the methods described up till now have failed catastrophically
because of short wavelength numerical instabilities that grow without bound.
To control these instabilities, we add some numerical diffusion. We do this
by adding a \( \nabla ^{4} \) term to the right hand sides of eqs. \ref{g dot},
\ref{K dot}, \ref{g_tt dot}, and \ref{Shift dot}. This technique, introduced
in \cite{Garp}, has the effect of spreading out short wavelength, poorly resolved
features, while leaving long wavelength, well resolved features alone. This
is exactly what is required --- we don't need an accurate evolution of short
wavelength features since they are poorly represented on a grid with finite
spacing. Thus, eq. \ref{g dot} becomes 
\begin{equation}
\label{g diffuse}
\frac{\partial \gamma _{ij}}{\partial t}=-2\alpha K_{ij}+\nabla _{i}\beta _{j}+\nabla _{j}\beta _{i}-q\left( \Delta x\right) ^{3}\nabla ^{4}\gamma _{ij},
\end{equation}
 where \( q \) is a small constant (we use \( q=0.09/R^{3}_{*} \), where \( R_{*} \)
is the radius of the neutron star). The \( \left( \Delta x\right) ^{3} \) factor
ensures that, as the resolution increases, the modified equation quickly converges
to the original continuum equation. We implement the \( \nabla ^{4} \) as \( \nabla ^{2}\left( \nabla ^{2}\right)  \),
and assume that \( \nabla ^{2}=0 \) on the boundaries. This is compatible with
the \( 1/r \) falloff, since \( \nabla ^{2}\left( 1/r\right) =0 \) on the
boundaries. The usual way to implement \( \nabla ^{2} \) is as centered second
derivatives along the coordinate axes (\( \nabla ^{2}=\partial ^{2}/\partial x^{2}+\partial ^{2}/\partial y^{2}+\partial ^{2}/\partial z^{2} \)).
Thus, a plot of the finite difference stencil in the \( x \)-\( y \) plane
looks like Fig. \ref{axes}.

This gives diffusion that acts primarily along the coordinate directions \( x \),
\( y \), and \( z \). Unfortunately, eq. \ref{Ricci} has mixed derivatives
of \( \gamma _{ij} \) (e.g., \( \partial ^{2}\gamma _{ij}/\partial x\partial y \)),
so points couple along diagonal directions. To fix this, we define new, diagonal
directions \( u=x+y \), \( v=x-y \), \( w=z \), and take derivatives along
these directions. Then the finite difference stencil looks like Fig. \ref{diagonal},
which couples along the diagonals in the \( x \)-\( y \) plane. We repeat
this for the \( x \)-\( z \) and \( y \)-\( z \) planes and average the
three different representations of \( \nabla ^{2} \). This gives us a stencil
that couples along the axes and diagonals. Finally, we apply this ``diagonal
diffusion'' to the \emph{new} time level. That is, ordinarily, without the
diffusion, the evolution equation is implemented as 
\[
\gamma _{\mathrm{new}}=\gamma _{\mathrm{old}}+\Delta t(RHS),\]
 where \( RHS \) is the right hand side of eq. \ref{g dot}. We add diffusion
by changing the update to 
\[
\gamma _{\mathrm{new}}=\gamma _{\mathrm{old}}+\Delta t\lbrace
RHS-q\Delta x^{3}\nabla ^{4}[ \gamma _{\mathrm{old}}+\Delta t\left( RHS\right)
]\rbrace .\]
 This is equivalent to first order to eq. \ref{g diffuse}. We implement it
this way because short wavelength instabilities can sometimes change sign at
each time step. Applying numerical diffusion to the old time step as in eq.
\ref{g diffuse} can end up being always a step out of sync with the instabilities,
adding when it should be subtracting. Implementing it in this way doesn't allow
the other parts of the equation (\( RHS \)) to interfere with the diffusion.

Adding explicit diffusion does slightly alter the equations away from the original
physical ones, although the change gets smaller with higher resolution. In that
sense, you can just consider it a slight addition to the error, getting smaller
with more resolution. However, the error is not random; it moves error in the
direction of too much diffusion. In an attempt to minimize this error, we tried
applying this diffusion only to the coordinate evolution equations (eqs. \ref{g_tt dot},
\ref{Shift dot}), because the coordinate evolution does not affect physical
quantities. Yet even when we explicitly enforced the constraints at each time
step, short wavelength spikes grew and destroyed the simulation.

This diffusion is very effective.  Ordinarily, second-order
Runge-Kutta is unconditionally unstable for most problems.  In this case,
instabilities seem to be dominated by other factors.  We have tried
alternative stepping alorithms, such as iterative Crank-Nicholson with
varying numbers of iterations, and, in every case, the evolution was
unstable without the diffusion, and stable with it.

\subsection{Computational Issues}

The entire code is written in C++ and runs on the Cornell Theory Center IBM
SP2 supercomputer. The code has also been ported to a Sun workstation, and is
freely available from the authors. The runs with \( 9^{3} \), \( 17^{3} \),
and \( 33^{3} \) points used 1 processor, the \( 65^{3} \) runs used 8 processors
and the \( 129^{3} \) runs used 64 processors. We used the Kelp libraries \cite{Kelp}
to split up the grid among multiple processors and simplify the message passing
between processors. The code scales from 1 processor to 64 processors with about
70\% parallel efficiency. All of the runs (except for the convergence tests,
which were very short) took 12-72 hours, taking 1000-1500 steps every 12 hours.
The hydrodynamics used about 80\% of the CPU time, the metric and coordinate
evolution used about 15\%, and everything else took up the remaining 5\%.

\section{Tests}

We performed two different types of tests: short-term tests to verify the correct
implementation of the equations, and long-term tests to verify the stability
and accuracy of the method.

\subsection{Short-Term Tests}

To test the hydrodynamic evolution and its ability to handle shocks, we ran
shock tubes as in \cite{Miller et. al.}. We set up a one dimensional shock
tube with \( \rho =10 \), \( P=13.3 \), \( v=0 \), on the left and \( \rho =1 \),
\( P=0.66\cdot 10^{-6} \), \( v=0 \) on the right. This problem has an analytic
solution \cite{Centrella 1984,Schneider 1992} which we can compare to our evolution.
Figure \ref{Shock} shows a comparison of the normalized values of the pressure,
density, and velocity, and we can see that our method reproduces the analytic
answer reasonably well.

To test the metric and coordinate evolution, we ran convergence tests with the
exterior metric of a black hole in full harmonic gauge \cite{Cook}. To test
all of the evolution terms and their coupling to each other all at once, we
ran convergence tests on static and boosted neutron stars. We performed our
neutron star tests with a \( \Gamma =\frac{5}{3} \), \( K=5.380\cdot 10^{9}\mathrm{cm}^{4}\mathrm{g}^{-2/3}\mathrm{s}^{-2} \)
polytrope with a central density \( \rho _{\mathrm{central}}=1\cdot 10^{15}\mathrm{g}\, \mathrm{cm}^{-3} \),
resulting in a mass \( M_{*}=1.35\cdot 10^{33}\mathrm{g} \), a Schwarzschild
radius \( R=12.7\mathrm{km} \), and thus a compaction \( M/R=0.08 \). The
maximum central density for this equation of state is about \( \rho _{\mathrm{max}}=3.79\cdot 10^{15}\mathrm{g}\, \mathrm{cm}^{-3} \),
corresponding to a maximum mass of \( 1.51\cdot 10^{33}\mathrm{g} \). Then
we converted the variables into the harmonic gauge \cite{Cook} which gives
us a harmonic radius of \( R_{*}=11.7\mathrm{km} \). We boosted the stars with
a variety of velocities \( \left[ \overrightarrow{v}=(0.3c,0.2c,0.1c),(0.9c,0,0),(0,0.5c,0)\right]  \)
using the prescription in \cite{Miller et. al.}. We did convergence tests over
both a cube entirely within the star and a cube entirely outside the star. We
offset the grid by \( 10^{-5}R_{*} \) in the \( x \), \( y \), and \( z \)
directions to avoid having to deal with spherical coordinate pathologies when
setting up the initial configuration.

When doing convergence tests, we first took one step on a \( 9^{3} \) grid,
two steps on a \( 17^{3} \) grid, and four steps on a \( 33^{3} \) grid. All
three grids spanned the same space, so each step doubled the resolution. Then we
compared the solution on the three grids against the analytic solution. Because
our differencing and time stepping are both second order, the error should decrease
as the square of the number of points (i.e., a factor of four with each step
in resolution).

However, this is not a perfect argument because there are higher order errors
contributing to the overall error. Fortunately, these errors decrease even faster
than the second order error. Therefore, if we plot the \( 9^{3} \) error, the
\( 17^{3} \) error multiplied by 4, and the \( 33^{3} \) error multiplied
by 16, then we should see the \( 17^{3} \) error closer to the \( 33^{3} \)
error than the \( 9^{3} \) error. 

Unfortunately, even this procedure is not foolproof. Sometimes the error can
go through zero, leading to cases where the \( 17^{3} \) error is closer to
the \( 9^{3} \) error than the \( 33^{3} \) error in isolated regions. Figure
\ref{Dconverge} shows a plot where this happens. Sometimes the higher order
error is still large, leading to cases where the high resolution error decreases
so quickly that the medium resolution error is no longer closer to it. As an
added difficulty, our outer boundaries aren't very good, so we couldn't use
the error near it. In practice, we had to throw away the two outermost points,
leaving only \( 5^{3}=125 \) points to compare. In addition, our solution of
the initial data for the neutron star is imperfect, so the center point was
often bad (For example, the only bad point for \( \gamma _{xy} \) is at the
center).

Finally, to get proper convergence, we had to turn off the numerical diffusion
terms, because they add unphysical, although small, diffusion, and also the
min-mod slope limiter in the hydrodynamics, because it enforces first order
convergence near maxima. Keeping these techniques allows longer evolutions while
sacrificing strict second-order convergence.

Even with all of these caveats, checking convergence was a powerful tool for
finding bugs. An error in how we implement an equation prevents the entire grid
from converging. Thus, a plot like Fig. \ref{K12converge}, which shows the
worst converging variable for that particular set of convergence tests, was
not worrisome. We can trace its defects back to a combination of what is wrong
in Fig. \ref{Dconverge} and problems at the center. When we can not do that,
the cause is a bug. We found many bugs this way. For example, we accidentally
implemented the \( -2\alpha K_{il}K^{l}_{j} \) term in eq. \ref{K dot} as
\( -2\alpha K_{il}K^{j}_{l} \), and the indexing error prevented convergence.
This sensitivity gives us a reasonable degree of confidence that we implemented
the correct equations, and implemented them without error.

\subsection{Long-Term Tests\label{long term tests}}

To test the ability of our code to handle long neutron star evolutions, we ran
the code with a single, stationary star. The idea is to see how well it holds
a static star in equilibrium. In this section and later we adopt the units \( R_{*}=11.7\mathrm{km}\equiv 1 \),
\( M_{*}=1.35\cdot 10^{33}\mathrm{gm}\equiv 1 \), and \( c\equiv 1 \). This
implies \( G=0.08 \), \( 0.039\mathrm{ms}=1 \), and the hydrodynamic timescale
\( \rho _{\mathrm{central}}^{-1/2}=4.49 \). We compute the norm of a variable
by averaging its square over the grid and then taking a square root.

The first set of tests evolved just the metric and coordinate terms, keeping
the matter terms fixed to their analytic values. Figure \ref{constraints} plots
the error in the energy constraint (eq. \ref{Energy}) vs. time for three cases.
It quickly drops down and levels off as the simulation settles into a stable
configuration. Ideally, we would like that configuration to be the analytic
one, but finite resolution, numerical diffusion, and the imperfect outer boundary
condition will change it. Figure \ref{energy_compare} plots a cross section
of the error for the three cases in the final state. The errors for the low
resolution, close and far boundaries cases are nearly identical. The only visible
difference is near \( x=2 \), where the close boundary has a little more error
because of the boundary. It is also clear that better resolution helps a lot,
drastically decreasing the error. This is probably because the biggest contributor
to the error here is the numerical diffusion terms. Their effect decreases as
the \emph{cube} of the grid spacing, which is much faster than the other errors,
which decrease as the \emph{square} of the grid spacing. The reason that the
constraint norm for the far boundary case is so much smaller than the close
boundary case is that the error is concentrated in the center near the star.
The far boundary case just has more points over which to average the error.
This is encouraging, because it means our simulation will improve as we increase
resolution (not always a given), and our boundary is probably adequate, even
relatively close in.

Figure \ref{sqrt_g_compare} plots a cross section of the relative error in
\( \sqrt{\det \gamma _{ij}} \). Even for the worst case, the errors are only
about \( 1\% \). The effect of the boundaries is much stronger than in Fig.
\ref{energy_compare}. The simulation does not stray far from the analytic solution,
giving us hope that numerical diffusion and the boundary conditions do not significantly
affect the results. 

The second set of tests evolved just the hydrodynamics, keeping the metric and
coordinate terms fixed to their analytic initial values. Figures \ref{diffuse_matter}
and \ref{diffuse_big_matter} show a cross section of \( T^{\mu }_{\mu } \)
versus time for two resolutions. The star rapidly diffuses out until it starts
to interact with the boundaries. Remember that the hydrodynamic terms do not
have any numerical diffusion. This diffusion doesn't actually cause a problem
until late times, when ``fingers'' appear, stretching from the star to the
boundary (Fig. \ref{finger}). These fingers quickly grow until they dominate
the simulation. The simulation still looks well behaved, if inaccurate, for
a long time before then (Fig. \ref{nofinger}).

The third set of tests evolved everything: hydrodynamics, metric, and coordinates.
The results are largely the same as the hydrodynamics-only results (Figs. \ref{diffuse_one},
\ref{diffuse_big_one}, and \ref{diffuse_wide_one}). We also ran a test with
far boundaries, and, as expected, we were able to evolve for significantly longer
times. Figure \ref{maxplot} shows a plot of the central density \( T^{\mu }_{\mu } \)
vs. time for the three cases. Eventually, the star again developed ``fingers'',
at which time the code became hopelessly inaccurate.

\section{Coalescence}

To show that the method will work for its intended purpose, simulating coalescing
stars, we simulated a coalescence with easily computed, although not astrophysical,
initial data. We took the equilibrium solution for the neutron star, replicated
it next to itself, and then boosted the stars with \( v=0.15c \) in opposite
directions as shown in Fig. \ref{overhead} (the Kepler frequency for two point
particles is \( v=0.19c \)). To get self-consistent gravitational initial data,
we solved the constraints with this matter source.

This is not astrophysically interesting initial data by any means. The equation
of state, and thus the size, mass, and internal structure of the stars, are
all wrong. Even if they were right, setting up the variables by placing boosted
solutions next to each other is definitely wrong. There is likely to be a large
amount of initial wave content, which can seriously affect the dynamics. We
discovered this to our chagrin when we tried starting the stars a little farther
apart. The initial wave content pushed the stars away from each other!

In addition, the resolution of these tests is really no better than the long
term tests in section \ref{long term tests}, which, as we saw, were not very
accurate. Even so, it serves as an important validation of the method.

To measure the gravitational waves, we adopt a scheme similar to that used in
\cite{Shibata cluster}. We can define the transverse traceless (TT) gauge wave
amplitudes \( h_{+} \) and \( h_{\times } \) by projecting out components
of the metric. Along the \( z \) axis, this becomes
\begin{eqnarray}
h_{+} & = & \frac{1}{2}\left( \gamma _{xx}-\gamma _{yy}\right) ,\label{h_plus} \\
h_{\times } & = & \gamma _{xy}.\label{h_cross} 
\end{eqnarray}
A more exact method would be to use gauge-invariant Moncrief variables \cite{Moncrief}
and integrate over spherical harmonics\cite{AbrahamsEvans}.
Considering the accuracy of the underlying
simulation, this would be overkill.

Figure \ref{merger} shows the gauge invariant trace \( T^{\mu }_{\mu } \)
vs. time for a low resolution run with the boundaries very far out at (\( -8 \),\( 8 \)).
The stars complete about two orbits before completely merging.

Figures \ref{h_plus_close} and \ref{h_cross_close} show the amplitude of \( h_{+} \)
and \( h_{\times } \) for three resolutions. As we increase the resolution,
the waveforms seem to converge, revealing finer and finer details. The atmosphere
was treated a little differently in these runs, with the floor value of \( D \)
set at \( 10^{-4}D_{\mathrm{central}} \) instead of \( 10^{-9}D_{\mathrm{central}} \).
This causes the evolution to go bad earlier, around \( t=60 \). The earlier
waveforms are not significantly affected, though. 

If we run some simulations with the boundaries farther out, the results are
not as encouraging. Plotting two resolutions of far boundary runs with the close
boundary, high resolution run (Figs. \ref{h_plus_far}, \ref{h_cross_far}),
we see that the initial amplitude and arrival time of the \( h_{\times } \)
wave is different. It arrives later, whereas if it were arriving from the \( z=0 \)
plane, where the stars are, it should arrive at an earlier time. This can not
be explained from differences in how the waves propagate from \( r=4R_{*} \)
to \( r=8R_{*} \). Figure \ref{h_cross_propagate} shows wave amplitudes for
the low resolution, far boundary run extracted at different radii. The \( h_{\times } \)
amplitudes match each other very well. 

Looking at the \( h_{+} \) wave, we see new structure around \( t=10-15 \)
(twin spikes) that is absent in the close boundary runs. This is probably just
propagation effects, for if we plot the wave amplitudes extracted at different
radii (Fig. \ref{h_plus_propagate}), we see that they match the structure much
more at \( z=4 \).

These differences in \( h_{\times } \) probably arise from differences in the
initial data. Remember that we solve the constraints to get the initial data.
We assume that the outer boundaries are correct, and change the interior metric
variables to fit. Thus, the different simulation domains will have varying amounts
of initial wave content. This also suggests, unfortunately, that the big, initial
peaks in \( h_{+} \) and \( h_{\times } \) may not be real. 

The stress energy tensor for weak waves in the TT gauge is \cite{Compact Objects}
\[
T_{\mu \nu }=\frac{1}{2}\left\langle h_{ij,\mu }h_{ij,\nu }\right\rangle ,\]
 where the \( \left\langle \ldots \right\rangle  \) symbol means an average
over several wavelengths. The total energy that passes through a thin shell
at radius \( r \) in time \( \Delta t \) is \( 4\pi T_{00}r^{2}c\Delta t \).
To estimate \( h_{ij,0} \), note that \( h_{+} \) and \( h_{\times } \) go
from \( \sim 10^{-2} \) to \( 0 \) over a time of \( \sim 10R_{*} \). We
extract the waves at \( 4R_{*} \), so the luminosity is 
\[
L\sim 4\pi \cdot \frac{1}{2}\frac{\left( h^{2}_{+}+h^{2}_{\times }\right) \left( 4R_{*}\right) ^{2}}{\left( 10R_{*}\right) ^{2}}=2\cdot 10^{-4}.\]

If we compute an estimate of the gravitational wave luminosity from the quadrupole
formula for point masses in circular orbits \cite[Section 36.6]{MTW}, 
\[
L=\frac{32}{5}\frac{\mu ^{3}M^{2}}{a^{5}}\sim 3\cdot 10^{-7},\]
 where \( \mu =m_{1}m_{2}/M \), \( M=m_{1}+m_{2} \), \( a=\mathrm{separation}=2R_{*} \).
The large differences in magnitude suggest that the waves may be generated by
something other the stars, such as the interaction between the right and left
sides of the grid (Fig. \ref{overhead}). The spacetimes are initially boosted
in opposite directions, so there is a discontinuity at \( x=0 \). This discontinuity
is smoothed somewhat, but not removed, when we solve the constraints.

\section{Future Directions}

There are a few things that must be done before the code will produce interesting
results. First, the code needs realistic initial data along with more realistic
equations of state. A number of groups \cite{Gourgoulhon,Wilson
irrotational,Uryu} have managed to construct reasonable
initial data. Getting these results and converting them into a usable format
for the dynamical code is, in principle, a straightforward task.

Second, the code needs better resolution. Simply increasing the number of points
is impractical, since the code already taxes the capabilities of current supercomputers.
Instead, the solution is probably adaptive mesh refinement (AMR) \cite{Berger}.
Massively parallel machines will \emph{still} be required, but at least then
the problem is possible. Unfortunately, AMR on massively parallel machines is
still a new endeavor, so libraries for making it straightforward to change a
normal, parallel code into parallel AMR code are still immature. AMR will also
allow us to move the boundaries out much farther, relieving some headaches there. 

These steps will allow the code to be applied to the most pressing reason for
these simulations: providing accurate theoretical waveforms for the new generation
of gravitational wave detectors coming on line. It will also give reasonably
accurate estimates of the amount of material ejected as possible \( r \)-process
elements. To understand \( \gamma  \)-ray bursts, on the other hand, will require
this and more. Implementing magneto-hydrodynamics will allow us to evaluate,
for example, the likelihood of the DRACO model \cite{DRACO}. Adding neutrino
generation and transport will account for an important energy source. Also,
neutrino effects may, as in supernovae, affect the dynamics \cite{Bethe}.

\section*{Acknowledgements}

We wish to thank Mark Scheel, Greg Cook and Larry Kidder for many useful
discussions. We would also like to thank Mark Miller for answering questions
about his numerical relativity code, Scott Baden for his help with getting Kelp
to work, Roger Lovelace for his help with hydrodynamics, and Thomas Baumgarte
for his help with outer boundary conditions. This work was supported in part
by NSF grants PHY 98-00737 and 99-00672, and
NASA Grant NAG5-7264 to Cornell University.

\begin{figure}
\caption{\label{axes}Finite difference stencil along axes. The circles denote which
points \protect\( \nabla ^{2}\protect \) samples in the \protect\( x\protect \)-\protect\( y\protect \)
plane. The points with X's do not affect the value of \protect\( \nabla ^{2}\protect \).}
\end{figure}
\begin{figure}
\caption{\label{diagonal}Finite difference stencil along diagonals.}
\end{figure}
\begin{figure}
\caption{\label{Shock} Normalized values of the density \protect\( \rho \protect \),
pressure \protect\( P\protect \), and velocity \protect\( v\protect \) for
a one-dimensional shock tube at \protect\( t=0.5\protect \) using 256 grid
points.}
\end{figure}
\begin{figure}
\caption{\label{Dconverge} Plots of the sign of the difference \protect\( \left| 4\cdot \left| 17^{3}\, \mathrm{error}\right| -\left| 9^{3}\, \mathrm{error}\right| \right| \protect \)
\protect\( -\left| 16\cdot \left| 33^{3}\, \mathrm{error}\right| -4\cdot \left| 17^{3}\, \mathrm{error}\right| \right| \protect \)
for the variable \protect\( D\protect \) for a neutron star with the boost
\protect\( \protect \overrightarrow{v}=\left( 0,0.5c,0\right) \protect \).
The boundaries are at \protect\( \pm 0.25R_{*}\protect \), but the outermost
2 points are discarded, so this grid is only 5 points on a side. When the difference
is positive, the normalized \protect\( 17^{3}\protect \) error is closer to
the normalized \protect\( 33^{3}\protect \) error than the \protect\( 9^{3}\protect \)
error, implying correct convergence. Thus, negative numbers (the dark areas)
indicate bad convergence. Note that the plot only goes negative in limited regions.
The 3rd order error goes through zero there, so the normalized errors cross
each other.}
\end{figure}
\begin{figure}
\caption{\label{K12converge} Plots as in Fig. \ref{Dconverge} but for the variable
\protect\( K_{yz}\protect \). For this set of convergence tests, this is the
variable that looks the worst.}
\end{figure}
\begin{figure}
\caption{\label{constraints} The norm of the error in eq. \ref{Energy} vs. time when
the matter terms are kept frozen for three cases: a low resolution run, a high
resolution run, and another low resolution run with the boundaries twice as
far out.}
\end{figure}
 
\begin{figure}
\caption{\label{energy_compare} The error in eq. \ref{Energy} for the relaxed state
on Fig. \ref{constraints}. The star is at the origin, and the star is spherically
symmetric. This plot is taken from the \protect\( x>0\protect \), \protect\( y=z=0\protect \)
line.}
\end{figure}
 
\begin{figure}
\caption{\label{sqrt_g_compare} A cross section of the relative error in \protect\( \sqrt{\det \gamma _{ij}}\protect \)
for the relaxed states of Fig. \ref{constraints}.}
\end{figure}
\begin{figure}
\caption{\label{diffuse_matter} A cross section of \protect\( T^{\mu }_{\mu }\protect \)
for a low resolution run (\protect\( 33^{3}\protect \) points) when only the
hydrodynamics are evolved. The star quickly spreads out, although the scale
is very small.}
\end{figure}
\begin{figure}
\caption{\label{diffuse_big_matter} A cross section of \protect\( T^{\mu }_{\mu }\protect \)
for a high resolution run (\protect\( 65^{3}\protect \) points) when only the
hydrodynamics are evolved. The star spreads out much more slowly than the low
resolution run of Fig. \ref{diffuse_matter}.}
\end{figure}
\begin{figure}
\caption{\label{finger} A cross section of \protect\( T^{\mu }_{\mu }\protect \) in
the equatorial plane taken at time \protect\( t=68.5\protect \). A finger has
developed along the \protect\( x=0\protect \) line.}
\end{figure}
\begin{figure}
\caption{\label{nofinger} A cross section of \protect\( T^{\mu }_{\mu }\protect \)
in the equatorial plane taken at time \protect\( t=50\protect \). There is
no evidence of a finger.}
\end{figure}
\begin{figure}
\caption{\label{diffuse_one} A cross section of \protect\( T^{\mu }_{\mu }\protect \)
for a low resolution run (\protect\( 33^{3}\protect \) points) when everything
is evolved. Note that it is very similar to Fig. \ref{diffuse_matter}.}
\end{figure}
\begin{figure}
\caption{\label{diffuse_big_one} A cross section of \protect\( T^{\mu }_{\mu }\protect \)
for a high resolution run (\protect\( 65^{3}\protect \) points) when everything
is evolved. Once again, very similar to the hydrodynamics-only result (Fig.
\ref{diffuse_big_matter}).}
\end{figure}
\begin{figure}
\caption{\label{diffuse_wide_one} A cross section of \protect\( T^{\mu }_{\mu }\protect \)
for a low resolution run, but with the boundaries twice as far as in Fig. \ref{diffuse_one}.
Note that it takes \emph{much} longer to reach the boundaries, resulting in
a much longer evolution.}
\end{figure}
\begin{figure}
\caption{\label{maxplot} The central density of the star for three different runs.
Note that the two low resolution cases track each other very well until the
close boundary case diverges dramatically.}
\end{figure}
\begin{figure}
\caption{\label{overhead}Overhead view of the initial state of the two neutron stars}
\end{figure}
\begin{figure}
\caption{\label{merger} \protect\( T^{\mu }_{\mu }\protect \) for coalescing stars.
The simulation spans (\protect\( -8\protect \),\protect\( 8\protect \)), but
we only plot (\protect\( -2\protect \),\protect\( 2\protect \)) to show the
details of the merger.}
\end{figure}
\begin{figure}
\caption{\label{h_plus_close} \protect\( h_{+}\protect \) (eq. \ref{h_plus}) for
three different resolutions extracted at the boundary \protect\( z=4\protect \). }
\end{figure}
\begin{figure}
\caption{\label{h_cross_close} \protect\( h_{\times }\protect \) (eq. \ref{h_cross})
for three different resolutions extracted at the boundary \protect\( z=4\protect \).}
\end{figure}
 
\begin{figure}
\caption{\label{h_plus_far} \protect\( h_{+}\protect \) (eq. \ref{h_plus}) for the
high resolution run of Fig. \ref{h_plus_close}, compared to low and medium
resolution runs with the boundaries twice as far out. The farther boundary wave
amplitudes are extracted at \protect\( z=8\protect \) and scaled to account
for the \protect\( 1/r\protect \) falloff. The structure differs markedly,
especially around \protect\( t=10-15\protect \).}
\end{figure}
\begin{figure}
\caption{\label{h_cross_far} \protect\( h_{\times }\protect \) (eq. \ref{h_cross})
for the high resolution run of Fig. \ref{h_cross_close}, compared to low and
medium resolution runs with the boundaries twice as far out. The farther boundary
wave amplitudes are extracted at \protect\( z=8\protect \) and scaled to account
for the \protect\( 1/r\protect \) falloff. The initial peak is significantly
stronger and arrives later.}
\end{figure}
\begin{figure}
\caption{\label{h_plus_propagate} \protect\( h_{+}\protect \) for a low resolution
run (\protect\( 65^{3}\protect \) points) with the boundaries at \protect\( \left( -8,8\right) \protect \).
The amplitudes are extracted at \protect\( z=4\protect \) and \protect\( z=8\protect \).
They do not match up, suggesting that the differences in Fig. \ref{h_plus_far}
may be due to propagation effects.}
\end{figure}
\begin{figure}
\caption{\label{h_cross_propagate} \protect\( h_{\times }\protect \) for a low resolution
run (\protect\( 65^{3}\protect \) points) with the boundaries at \protect\( \left( -8,8\right) \protect \).
The amplitudes are extracted at \protect\( z=4\protect \) and \protect\( z=8\protect \).
They match up reasonably well, especially the large initial peak.}
\end{figure}

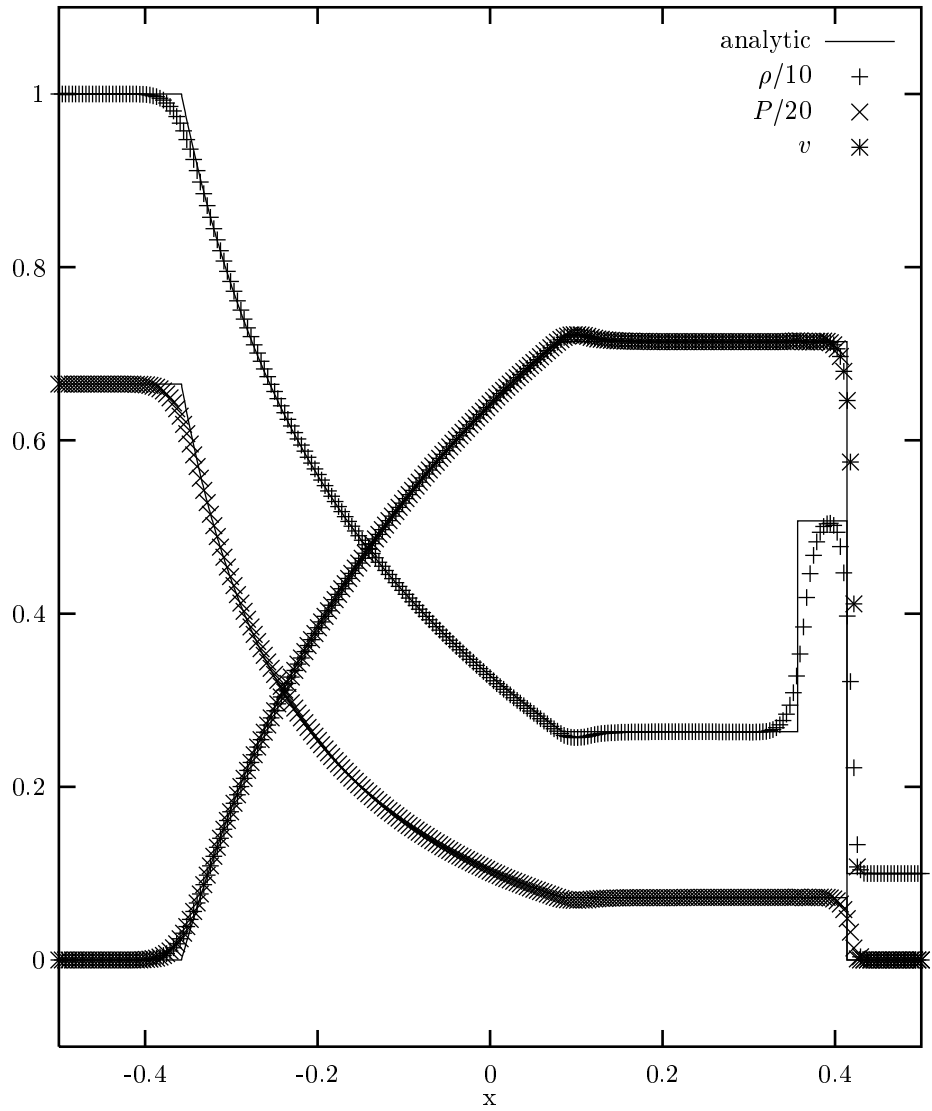
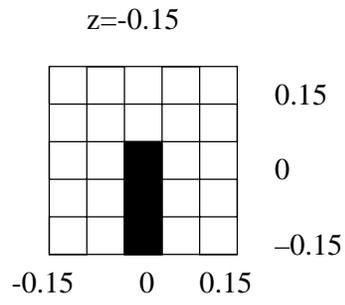
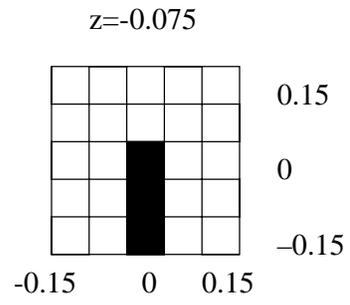
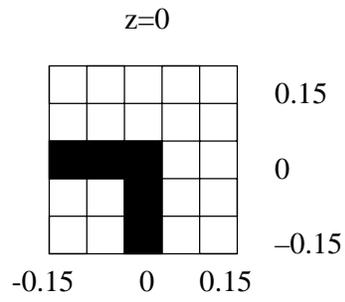
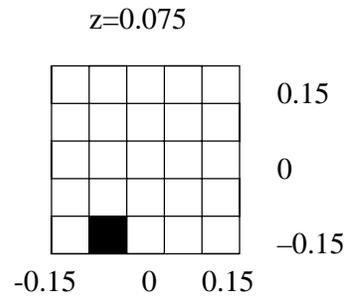
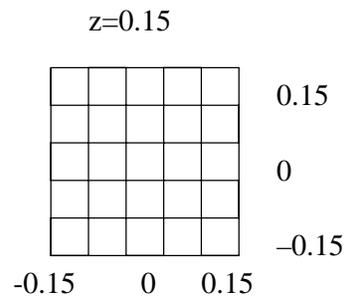
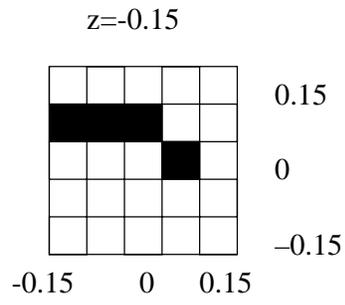
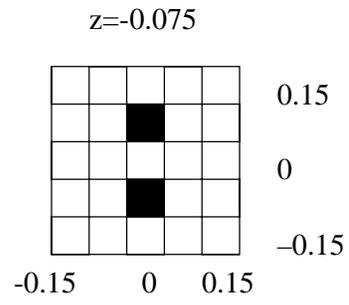
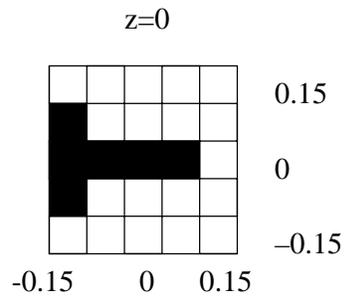
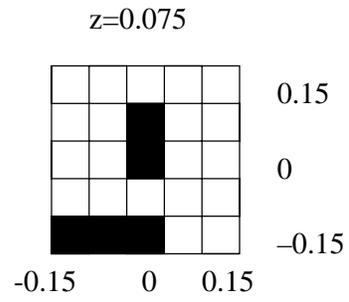
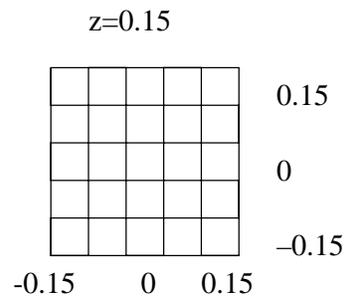
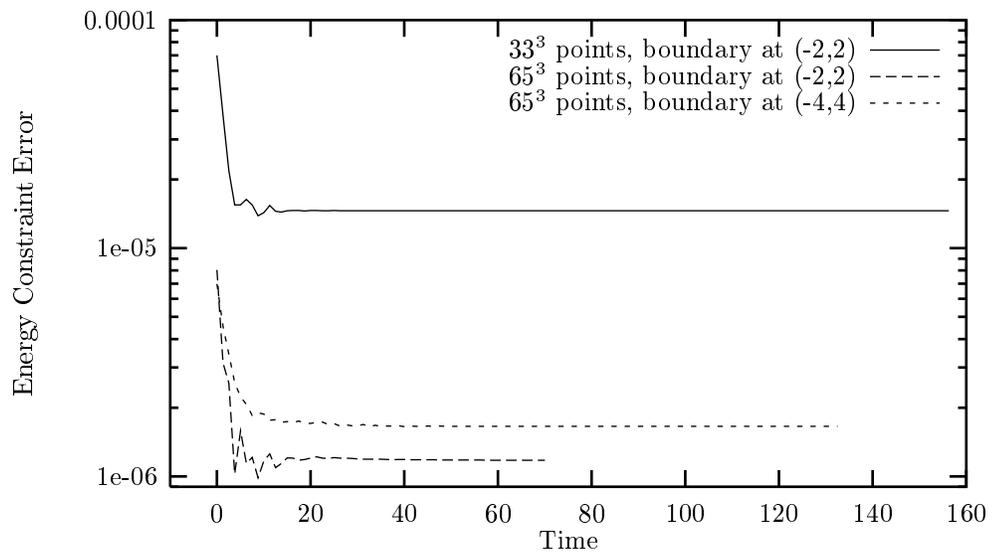
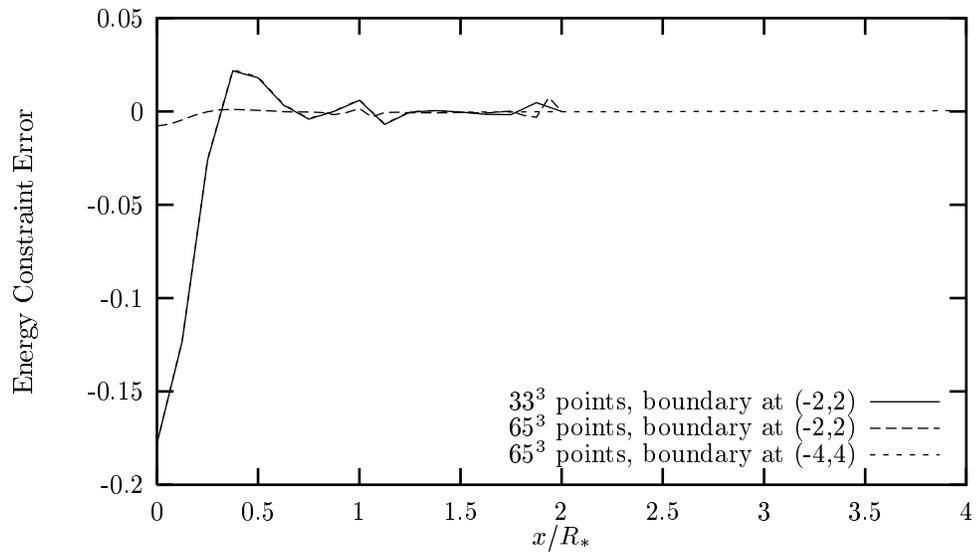
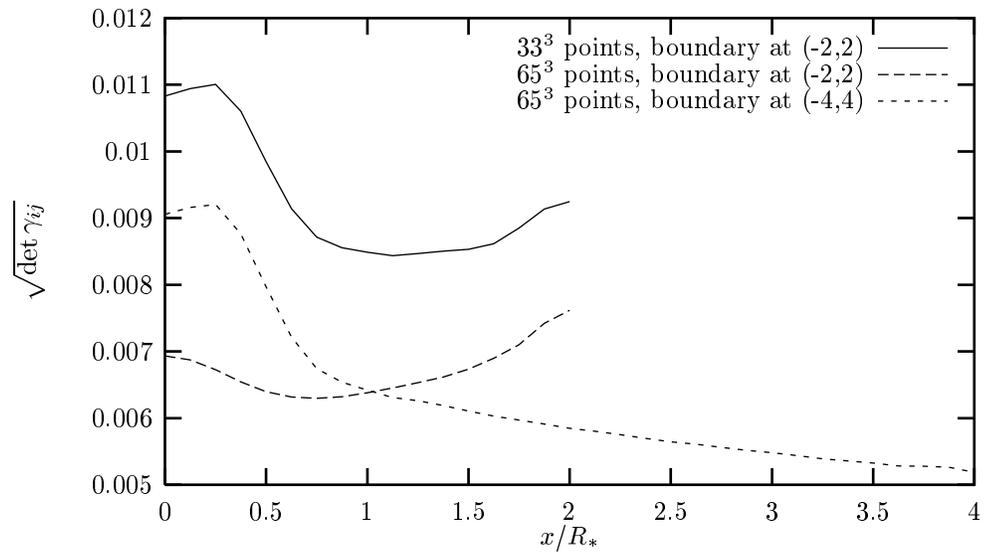
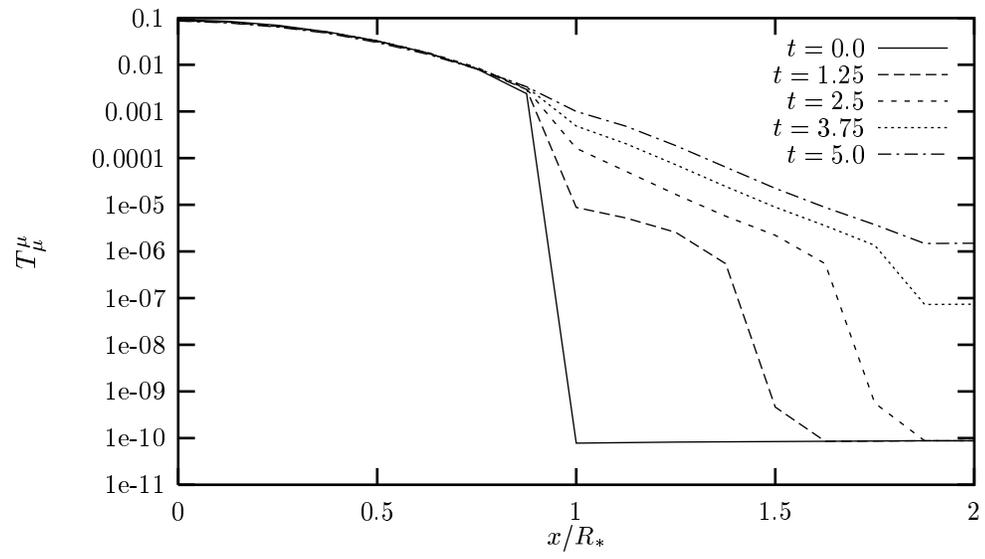
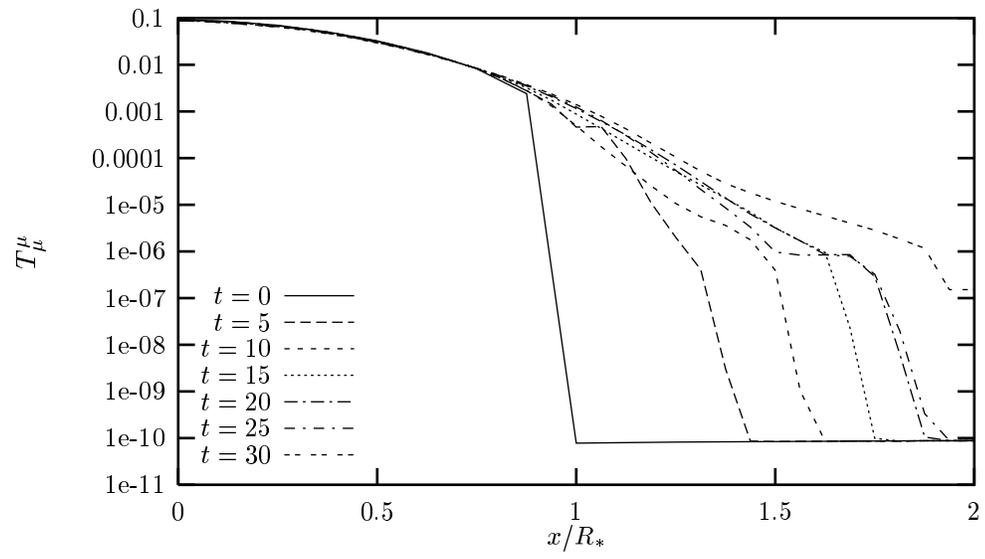
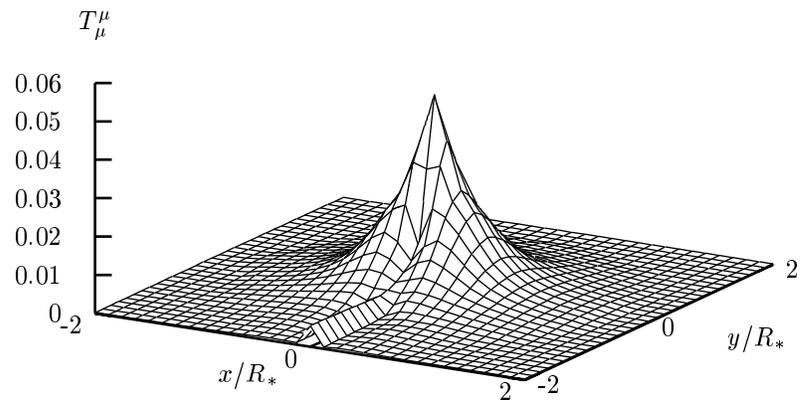
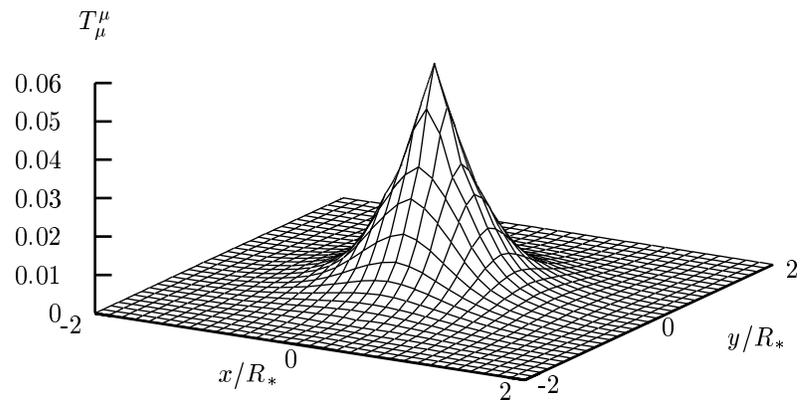
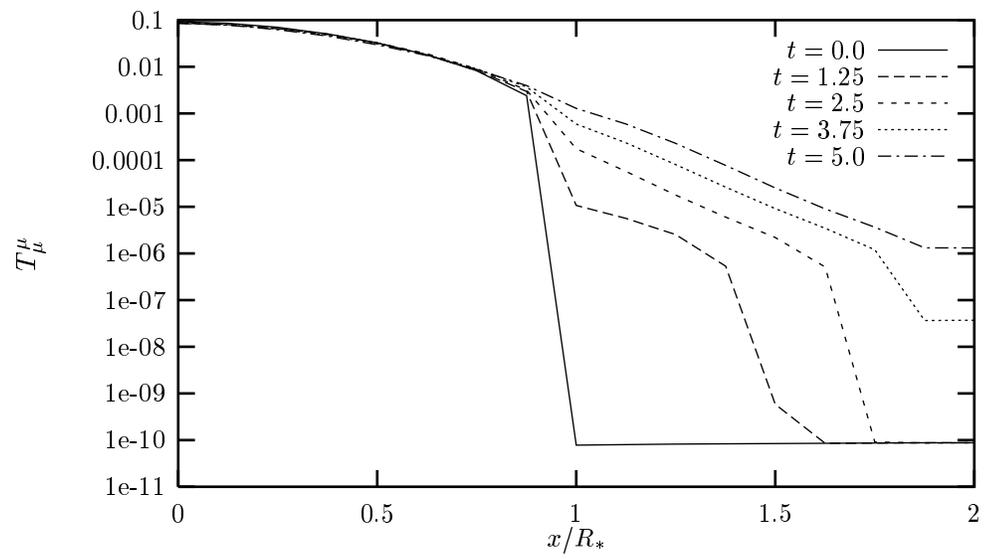
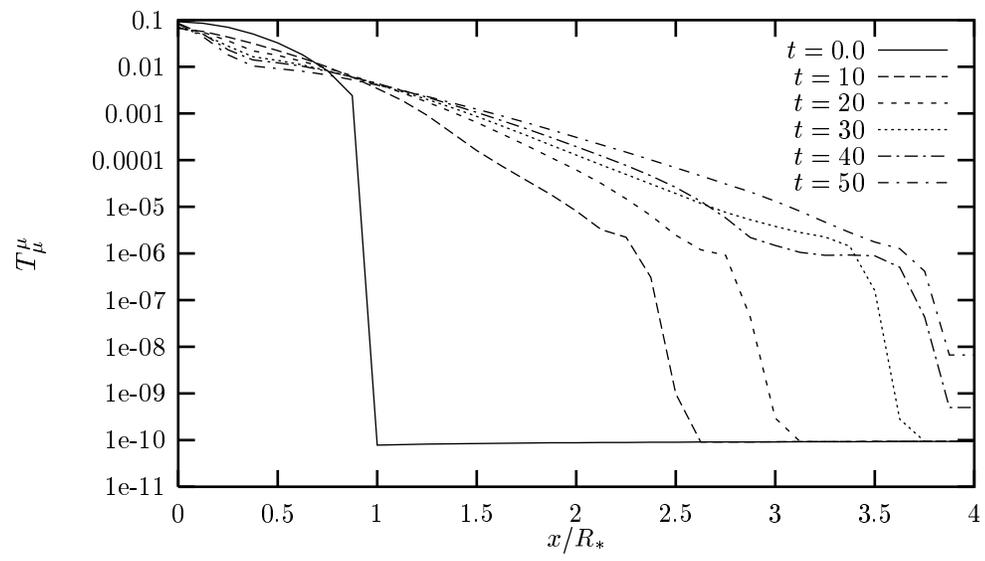
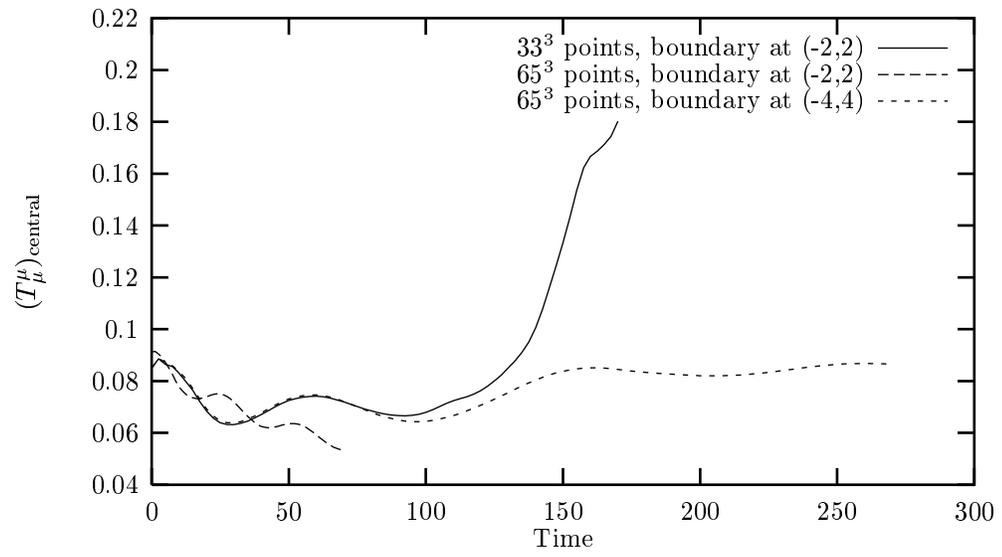
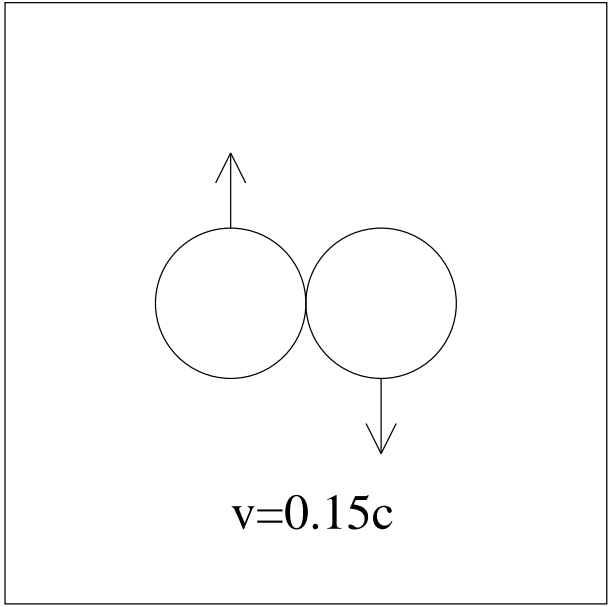
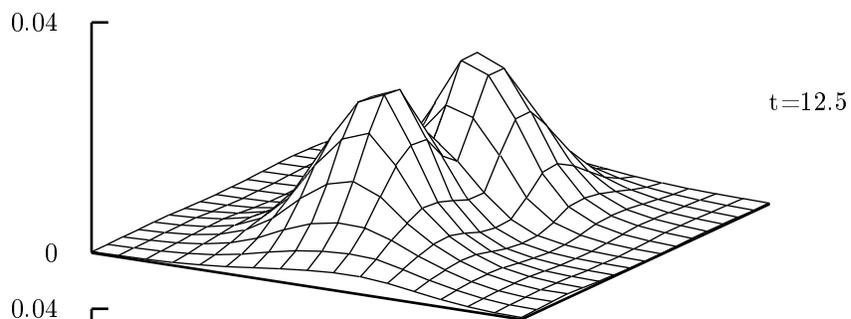
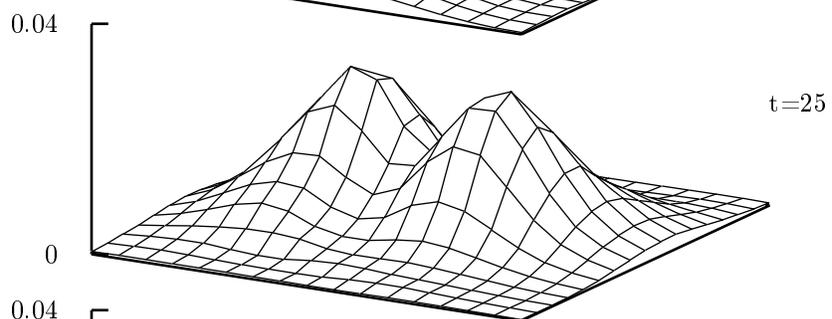
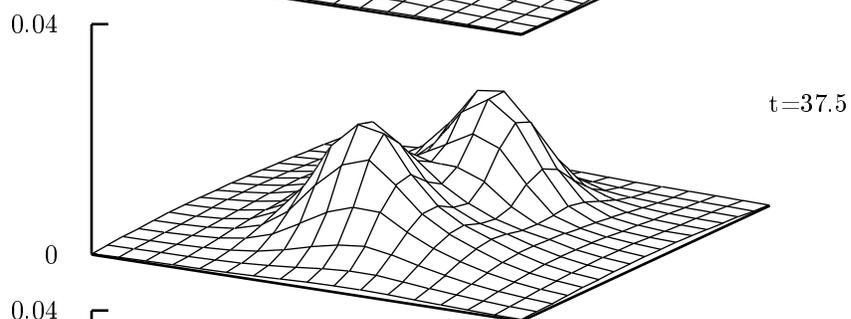
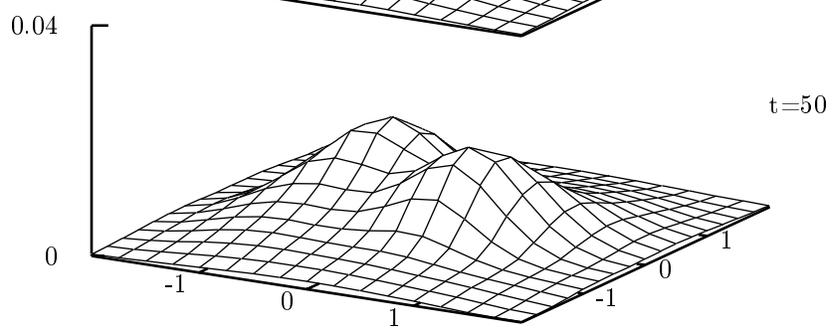
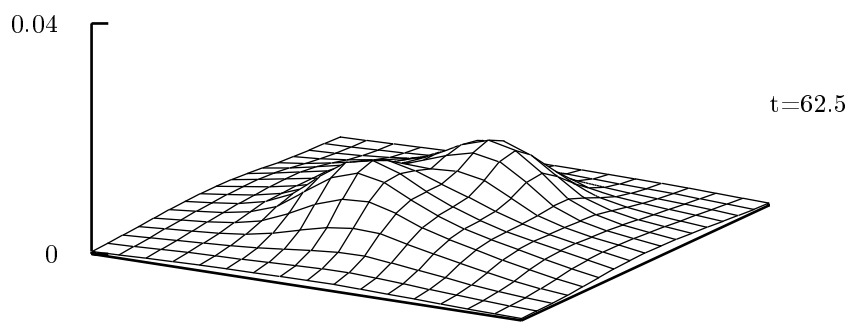
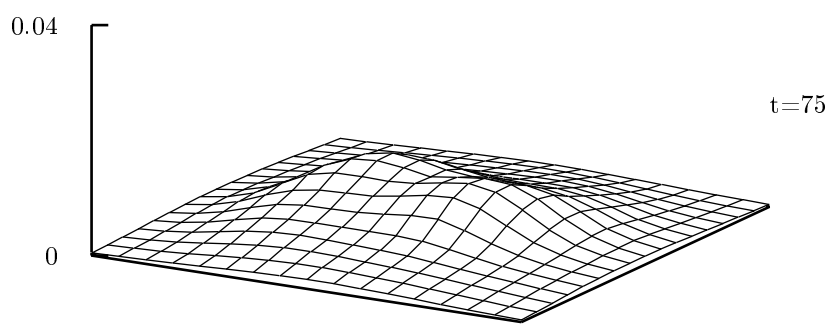
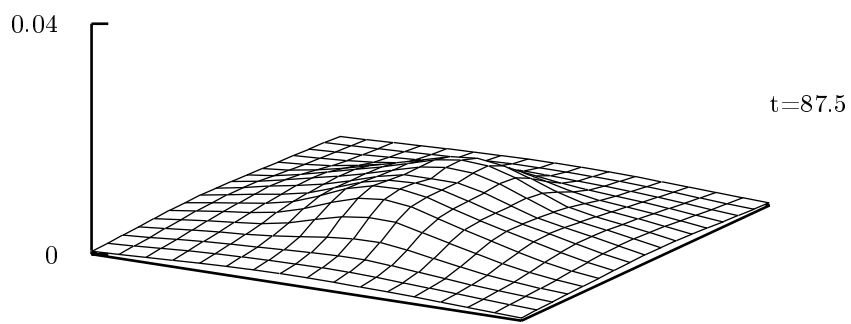
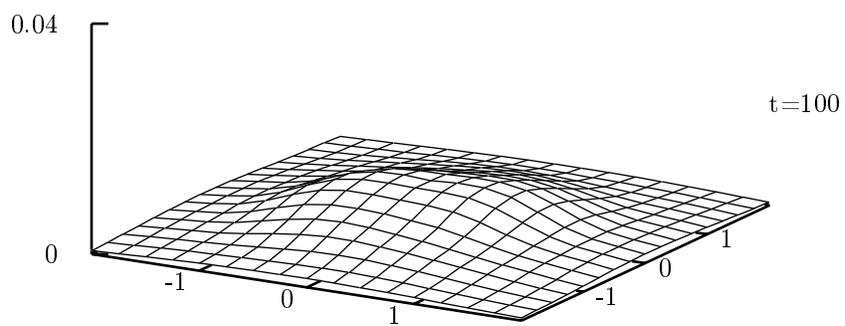
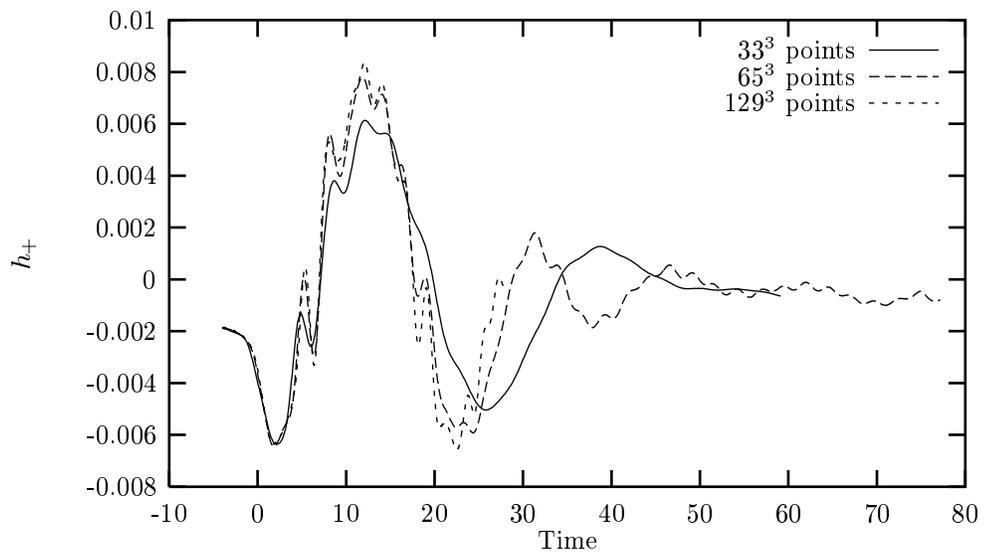
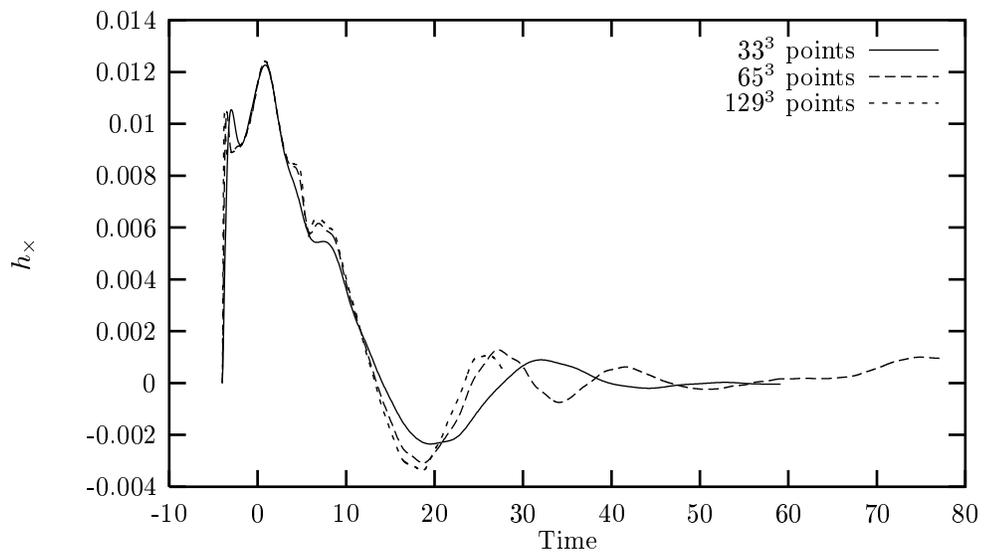
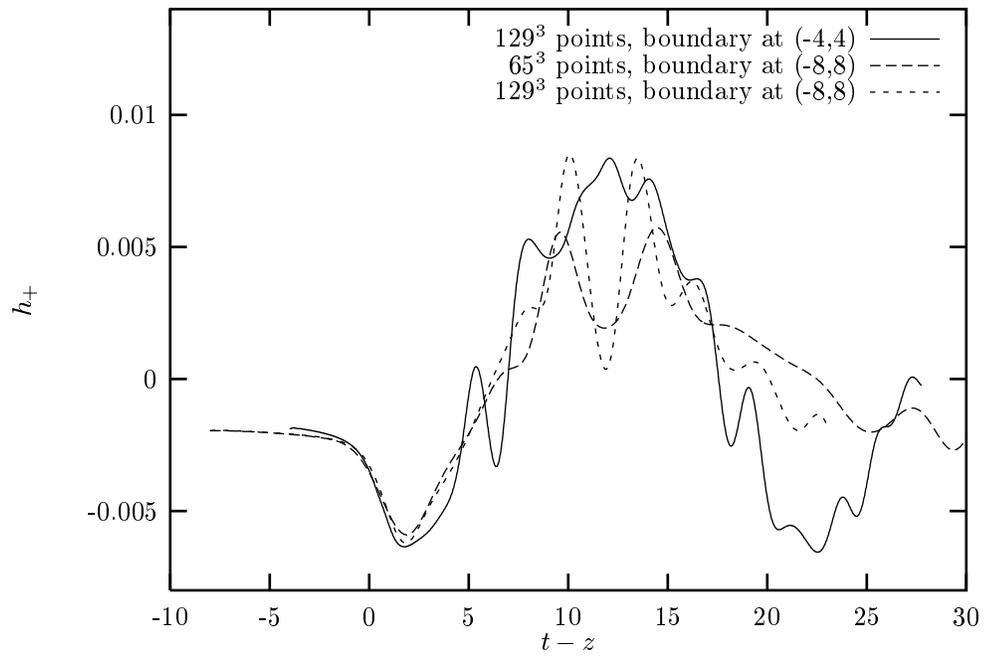
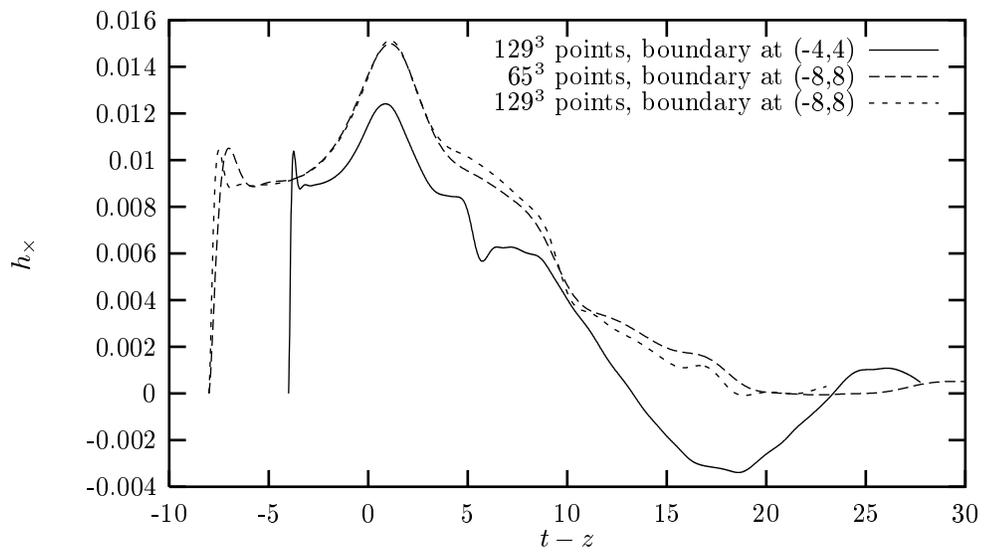
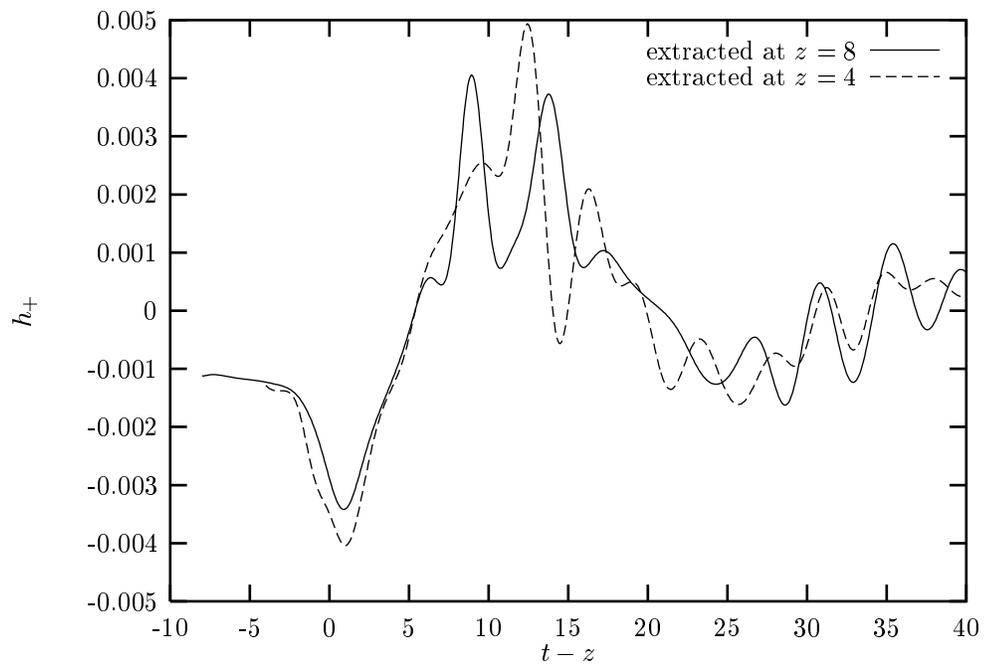
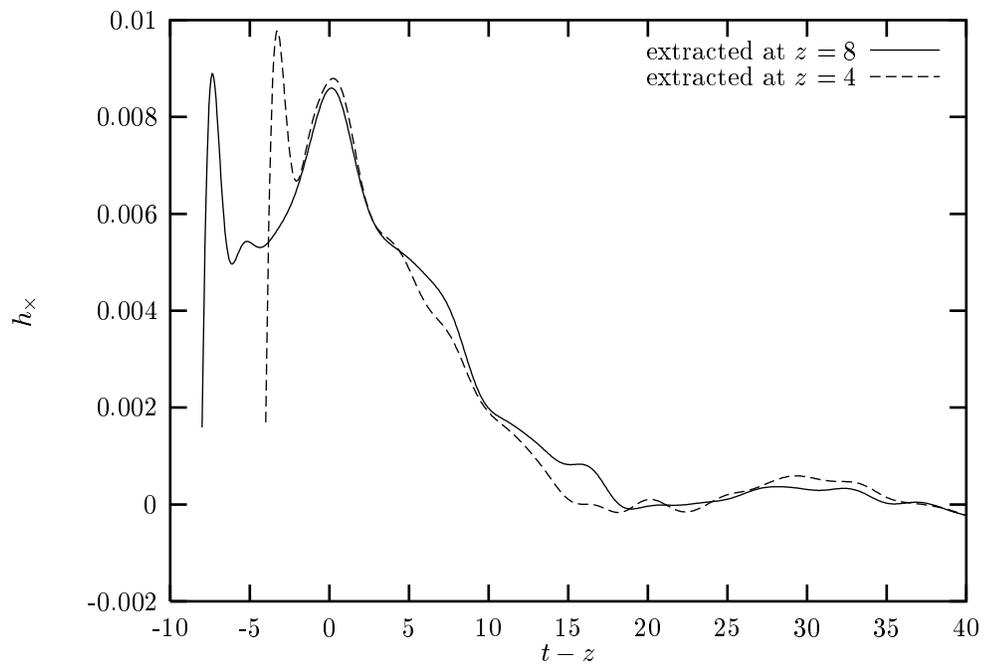
\end{document}